\documentclass[conference]{IEEEtran}
\IEEEoverridecommandlockouts
\usepackage{cite}
\usepackage{amsmath,amssymb,amsfonts}
\usepackage{graphicx}
\usepackage{textcomp}
\usepackage{xcolor}

\usepackage[bookmarks=false,hidelinks]{hyperref}
\usepackage{algorithmicx}
\usepackage{algpseudocode}

\ifCLASSOPTIONcompsoc \usepackage[caption=false,font=normalsize,labelfont=sf,textfont=sf]{subfig}
\else
\usepackage[caption=false,font=footnotesize]{subfig}
\fi

\DeclareMathAlphabet{\mathbf}{OT1}{cmss}{bx}{n}

\def\BibTeX{{\rm B\kern-.05em{\sc i\kern-.025em b}\kern-.08em
    T\kern-.1667em\lower.7ex\hbox{E}\kern-.125emX}}
\begin{document}

\title{Experience Management in Multi-player Games\\
}

\author{
\IEEEauthorblockN{Jichen Zhu}
\IEEEauthorblockA{\textit{Drexel University}\\
Philadelphia, PA, USA \\
jichen@drexel.edu}
\and
\IEEEauthorblockN{Santiago Onta{\~n}{\'o}n*\thanks{*Currently at Google}}
\IEEEauthorblockA{\textit{Drexel University}\\
Philadelphia, PA, USA  \\
santi@drexel.edu}
}

\IEEEpubid{\begin{minipage}{\textwidth}\ \\[12pt]
978-1-7281-1884-0/19/\$31.00 \copyright 2019 IEEE
\end{minipage}}

\maketitle

\begin{abstract}
Experience Management studies AI systems that automatically adapt interactive experiences such as games to tailor to specific players and to fulfill design goals. Although it has been explored for several decades, existing work in experience management has mostly focused on single-player experiences. This paper is a first attempt at identifying the main challenges to expand EM to multi-player/multi-user games or experiences. We also make connections to related areas where solutions for similar problems have been proposed (especially group recommender systems) and discusses the potential impact and applications of multi-player EM.
\end{abstract}


\section{Introduction}\label{sec:intro}

Experience Management studies AI systems that automatically adapt interactive experiences such as computer games to better serve specific users and to fulfill specific design goals~\cite{weyhrauch1997guiding,bates1992virtual,riedl2008dynamic,thue2018toward}. For example, experience management techniques have been designed to adapt interactive game experiences to follow a desired story arc~\cite{weyhrauch1997guiding}, or to dynamically adjust the difficulty of a game~\cite{hunicke2005case}. Experience management techniques have been explored for several decades. Its effectiveness has been demonstrated in commercial deployment in games such as {\em Left 4 Dead 2}. So far, most work in this area focused on single-player interactive experiences. 

Expanding {\em experience managers} (EMs) to multi-player scenarios is not trivial. As a research community, we lack the knowledge of how to effectively combine multiple player models and adapt the game accordingly to provide the desired gameplay experience to multiple players with different needs and preferences at the same time. Nor do we fully understand how to define and measure the desired collective experience.


Despite its challenges, the potential benefit of being able to adapt to multiple players simultaneously is high. It would enable the application of EM techniques to a large variety of scenarios such as massive multi-player online games (MMOs) as well as to serious games or training simulations that target groups of users, or the emerging area of social exergames or games for health~\cite{caro2018understanding}. It would also facilitate shared play experiences between players with different skill levels. For instance, grandparents play with grandchildren.

This paper analyzes the main challenges and provide a road map to research multi-player EMs, namely: {\em player model aggregation}, {\em dynamic player grouping/aggregation over time}, {\em integration with procedural content generation}, or {\em evaluation methods}. Additionally, we make connections to related areas where solutions for similar problems have been proposed (such as group recommender systems~\cite{jameson2004more}).

The remainder of this paper is organized as follows. We begin by providing background on EM and user/player modeling. We follow by elaborating on the potential impact of addressing the problem of multi-player experience management before moving on to identifying a set of open challenges in this area, and propose venues for future work.

\section{Background}\label{sec:background}


\subsection{Experience Management}

\begin{figure}[tb]
    \centering
    \includegraphics[width=0.8\columnwidth]{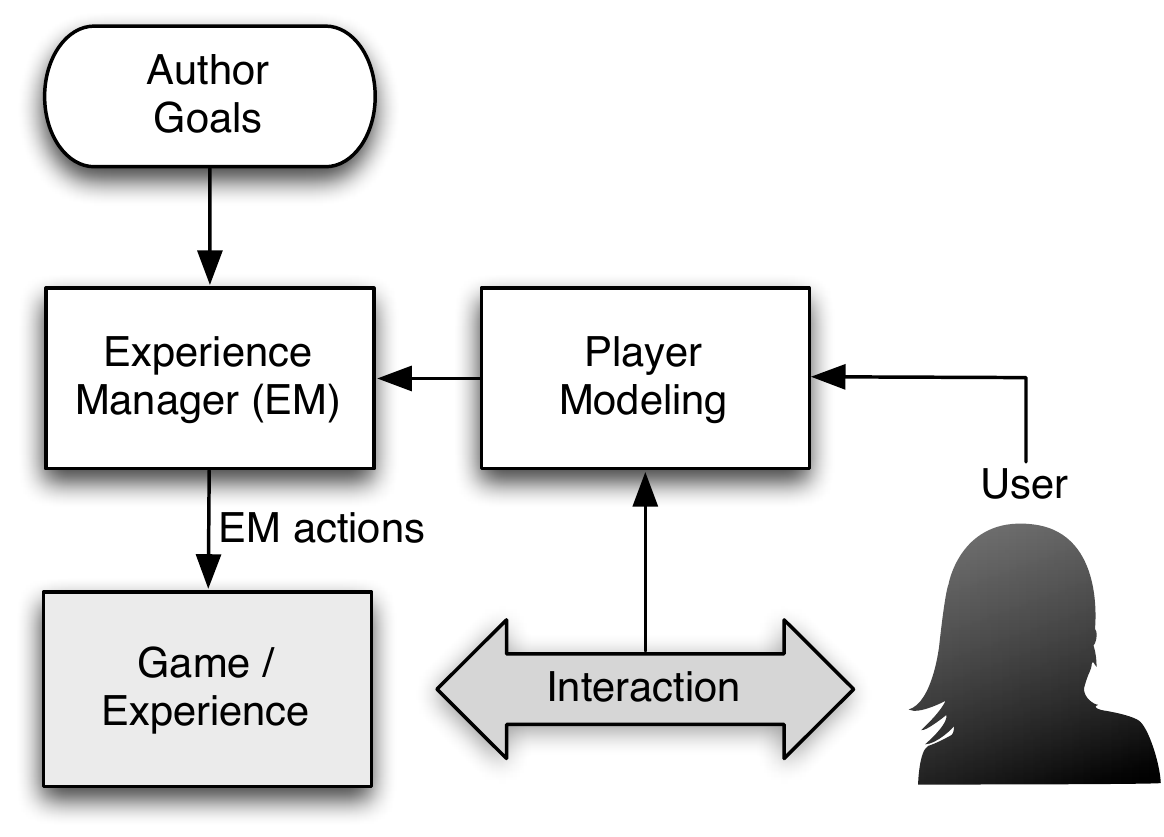}
    \caption{The main control loop of a standard experience manager, which observes the interaction of a player with a game, creates a player model based on this information and then considers the player model and/or author-defined goals to adapt the game via EM actions.
	}
    \label{fig:em}
\end{figure}

An {\em Experience Manager} (EM) 
is an AI component that oversees how one or more players interact with a game or other types of interactive experience, and adapt the game according to some predefined criteria (which we will call the {\em objective function}). As illustrated in Figure \ref{fig:em}, the EM observes the interaction between the player(s) and the game (i.e., game state and actions performed by the player), and uses a certain decision mechanism to determine how to adapt games to maximize or satisfy the objective function. These adaptations are usually formalized in the literature with the concept of ``EM actions'' (the set of changes the EM can do to the game). The objective function could be an author-defined set of goals (e.g., trying to maintain the tension of the story as close as possible to an Aristotelian arc), it could be a function that depends on the current player(s) (e.g., lead the player toward an ending that she will enjoy the best), or it could be a combination of both. When the function depends on the current player, the EM usually needs to build and maintain a {\em player model} to help evaluate the objective function. For example, if the goal of the EM is to adjust the difficulty level, a player model that reflects the skill level of the current player can be used. Thus, experience management is closely related to the field of {\em player modeling}.

Experience management research is not limited to entertainment domains but draws on decades of research on how to adapt and enhance interactive learning environments, including intelligent tutoring systems\cite{anderson1985intelligent, anderson1995cognitive, koedinger2007exploring}, pedagogical agents~\cite{johnson2000animated, moreno2001case, choi2006cognitive}, and cognitive science/AI-based learning aids~\cite{kolodner2000theory}. Specifically, the idea of an EM came out of research in drama management (DM)~\cite{weyhrauch1997guiding, mateas2002behavior, riedl2008dynamic}, a particular case of EM that focuses on balancing player agency and the authorial intent to maintain the narrative quality in games~\cite{aylett2003towards, cavazza2002character}. Initially proposed by Laurel~\cite{laurel2013computers} and Bates~\cite{bates1992virtual}, a drama manager is an AI system that functions as a virtual ``director'' who (re)arranges how the story should unfold while the player interacts with a game. Experience management extends the core concept of DM to domains beyond interactive drama, such as educational games~\cite{aylett2005fearnot, rowe2011integrating, marsella2000interactive, lee2011director, sabourin2012early}. For example, an EM could be designed to optimize an objective function based on pedagogical goals. A wide variety of AI techniques have been explored in the context of EM, such as search-based techniques~\cite{weyhrauch1997guiding, nelson2005search, sharma2010drama}, reinforcement learning~\cite{nelson2006declarative, wang2016decomposing}, or automated planning~\cite{magerko2004ai, young1999notes, young2004architecture}, among others. Despite this large amount of work, research in experience management remains mostly limited to single player experiences.

An exception to this is the work on difficulty balancing in multi-player games of Prendinger et al.~\cite{prendinger2016extending}. They presented an approach based on distributed constraint optimization to adjust the difficulty of scenarios according to the skill levels of different players. Another exception is the work of Riedl et al.~\cite{riedl2011robust}, who study the problem of allowing human authors to create stories for multi-player experiences, and how to design an automated narrative system that detects when players are deviating from the authored storyline, and makes changes to correct it. To extend this limited set of existing work, several key open research challenges need to be addressed, which we elaborate on in this paper.

\subsection{Player Modeling}

In the context of game environments, a player model is an abstracted description of a player capturing certain properties of interest such as preferences, strategies, strengths or skills~\cite{van2003local}. Player modeling has been used to model aspects such as player behavior (using unsupervised learning techniques~\cite{Drachen2009,valls2015exploring}, or for behavior prediction using supervised learning~\cite{holmgaard2014evolving,harrison2011using}), to identify player types~\cite{bartle1996hearts,heeter2011beyond}, to detect fun/boredom/engagement/frustration/challenge~\cite{pedersen2009modeling}, user preferences~\cite{thue2008passage,sharma2010drama}, skill level~\cite{missura2009player,jennings2010polymorph,zook2012temporal} and even player goals~\cite{ha2011goal}. For an overview on player modeling, the reader is referred to existing surveys of the area~\cite{smith2011inclusive,Machado2011}, \cite[Chapter~5]{yannakakis2018artificial}. As with experience management, most player modeling approaches focus on modeling individual players. 


{\bf Group Player Modeling.} As we will discuss below, when working in multiplayer EM settings, {\em group player modeling} might be necessary. By group player modeling, 
we refer to modeling ``groups of players'', and not ``players in a group''. For example, it might be necessary to predict whether a particular group of players would get along; or what would the behavior of the group as a whole be.
Notice that, although there are many approaches that apply player modeling to understand user populations or ``classess'' of players (by creating player types, or general trends of different groups of players), their goal is not to model a team or group of players currently playing the game, but to make predictions about individual players by modeling overall trends of larger populations. Thus,  even if it uses data from multi-player games, such as the work on player modeling in MOOs~\cite{shen2012inferring}, or are based on large amount of player data~\cite{weber2011modeling}, we do not classify this work under the ``group player modeling'' category.



\section{Multi-player EM Applications}\label{sec:impact}

Classic EM has expanded the capacity of computer games by tailoring the experience for a single player. Multi-player EM can make similar advancement. This section highlights some application domains that can benefit significantly. 

\begin{itemize}
    \item {\em Adaptive MMOs}: As massive multi-player online (MMO) games remain a popular genre, online games that allow for multi-player gameplay and support social interactions between players have become the norm. While deploying EM algorithms to games with several million players is perhaps beyond the reach of the state of the art, adaptive multi-player online games are an interesting application. As an illustration, consider the following example. Popular MMOs receive regular patches from the developers adding quests and continuing the storyline. This could be complemented with a story-driven EM constantly monitoring different player groups and manipulating the story to automatically trigger plot points to advance the story based on these player group models and vaguely defined story directions by the game designers. This would achieve more customized storylines, beyond the one-size-fits-all model. This EM could monitor and implement changes at the individual player level, but also at higher-levels such as factions of players or even the whole game, taking into account the actions of all the players involved. This is related to the idea of {\em AI as a game producer} put forward by Zook and Riedl~\cite{riedl2013ai}. 
    
    \item {\em Multi-user training and educational applications}: while training and educational games have received significant attention in the literature of EM and player modeling~\cite{valls2015exploring, rowe2010modeling}, an overwhelming majority focuses on single player experiences. However, learning science has shown that learning in groups can be beneficial \cite{slavin1980cooperative,johnson2008active}. Muti-player EMs can better support learning in groups. For example, active learning theory \cite{johnson2008active} suggests to team up students in a small, heterogeneous (in terms of perspectives and skills) group with  a group goal that can only be accomplished by the group together. In an educational game designed based on this theory, a multi-player EM is particularly relevant. As a heterogeneous group means that each student is meaningfully different from one another, the same player model or adaptation strategy may not fit everyone. Notice that a single player model captures the ``average of all the players''. Furthermore, having a shared goal that requires everyone's contribution means that the game needs to coordinate with different players so that they all stay engaged. In this case, having individual models for each learner and a centralized on-line agent to dynamically coordinate every players' learning needs and preferences can be more effective than using the same model and EM actions for everyone. 
    
    
    \item {\em Adaptive Games for Health}: Similar to training and educational games, adapting the gameplay experience for a group of people can benefit those who use games for health. Take exegames, a genre of video games that combines gameplay and physical activity, as an example. Popular games in this genre include {\em Pok\'{e}mon Go}. Although adaptive technology has been used in exergames (e.g., automatically adjusting difficulty, or daily exercise goals), how to engage players for a period of time that is long enough for potential behavioral change is still an open problem. Researchers have increasingly looked into multi-player exergames and use the social interaction between players as an additional motivation for physical activities.  Multi-player EM can help with the player retention issue as it helps to keep exergames more engaging for individual players and the team as a whole.
    
\end{itemize}


Finally, notice that this is not an exhaustive list of potential applications. Many others such as any multiplayer game with AI opponents, or even computer-assisted table-top games, where the AI is attempting to balance the game to even out player skills, or any other measure of interest.

\section{Open Challenges in Multi-player EM}

In order to realize the potential applications highlighted above, there are a number of open challenges to be addressed. This section briefly lays out the road map of multi-player EM.

\subsection{Player Model Aggregation}\label{sec:aggregation}


Traditional EM approaches use models of a single player to adapt the game. These models capture aspects of a player that are of interest to the game, such as her interest \cite{sharma2010drama,thue2008passage} and play style \cite{valls2015exploring}. If, for example, a system like PaSSAGE~\cite{thue2008passage} determines the current player prefers social interactions rather than combat interactions, it can alter the game and cater to that preference. However, when there are multiple players, their preferences may not be aligned. How to aggregate the content of individual player models (e.g., gameplay preferences, learning styles) is an open problem when we extend EM to multi-player. Currently, we lack theories and understanding of 1) how to aggregate multiple models, 2) how to dynamically group players represented by multiple models, 3) how to adapt the gameplay experience based on multiple/aggregated models, 4) how to evaluate the effectiveness of multi-player EMs. 

To ground the discussion, we use player preference as an example. However, the crux of the following discussion is applicable to other aspects of players we might want to model.  

Preference aggregation is an interdisciplinary field of study that has looked at this problem from different perspectives, such as group recommender systems~\cite{masthoff2011group}, or multiagent systems~\cite{pini2011incompleteness}. However, so far, preference aggregation has not been sufficiently explored in the area of player modeling in the context of games.
For player model aggregation, a particularly relevant area to build upon is {\em group recommender systems}, which are designed to provide recommendations to groups of users. Jameson~\cite{jameson2004more} identified four key challenges that are unique to group recommender systems, namely: preference elicitation (i.e., should users enter their preference individually or collaboratively), preference aggregation, explanation of recommendation suitability to the different group members, and supporting joint decision making by the group to reach a final recommendation. 
Early work to address the preference aggregation problem includes systems like  MusicFX~\cite{mccarthy1998musicfx} and PolyLens~\cite{o2001polylens}. MusicFX focuses on adjusting the music selection for group workouts based on pre-specified preferences, in the form of music channels, of all the users beforehand. At any given time, the system adds up the preferences of each user. One of the top $m$ most preferred channels is selected randomly ($m$ to increase variety). PolyLens performs group recommendation using collaborative filtering. It generates a recommendation list for each user and then merges these lists. The items in the merged list are sorted according to a ``social value function'' that represents an aggregation of the preferences of the members of the group. 
The key idea behind these two systems is that of {\em adapting to group preferences by merging or averaging the individual preferences}. 
Another, less common idea is that of negative preferences~\cite{chao2005adaptive}: detect what users do not like, and avoid those (which is often easier than detecting what they like).

Although those techniques focus on recommending individual items out of a predefined set, and thus might not directly apply to many domains where we want to apply EM, they provide a starting point. For example, even if the usual collaborative filtering algorithm used in recommender systems might have to be replaced by something else, narrative-based EMs could adapt these ideas to use them to ``recommend the next plot points'' in a multi-player story.



Finally, notice that the idea of player model aggregation is fundamentally different from that of using clustering for player modeling. Clustering approaches aims to find player types/classes and group players by their similarity in some features of interest. In other words, clusters reflect the average properties of groups of players that are similar in those features of interest. By contrast, player model aggregation focuses on aggregating preferences of players that happen to be playing together in the game, and might bear no resemblance in their behavior. These players might have never been clustered together using clustering.

\subsection{Dynamic Multi-player EM over Time}\label{sec:fairness}


Gameplay experiences change over time and thus how to best adapt to a group of players may need to shift accordingly. How to aggregate multiple player models differently over time is one of the least studied research areas. 

\subsubsection{Dynamic Aggregation Mechanics}
So far, most existing work uses static aggregation mechanics to combine the preferences of multiple players. For example, {\O}lsted et al. \cite{olsted2015interactive} use majority vote to evolve game levels for a multi-player shooter game. The limitation of this approach is that if a player's preferences are always overlooked, she is more likely to quit and negatively impact the gameplay for other players.  Consider a situation where the majority of players in a group have already mastered the basics, but one single player (or a small group) is still struggling. Under a static majority-vote aggregation mechanic, this minority group would be constantly underserved. An EM framework based around a majority-voting or mean-based aggregation of player models, therefore, will inevitably lead to {\em unfairness over time}. 


With a few exceptions~\cite{baccigalupo2007case}, group recommender systems have not explored this area extensively either. For example, if there is a small group of users with a very different taste in MusicFX, their preferences will not outweigh those of the larger group, and thus, their dissatisfaction will grow over time. One exception is a group music recommender system called {\em Poolcasting}~\cite{baccigalupo2007case}. In its aggregation function of individual preferences, the more a user is unsatisfied with the recent songs, the more her preferences will influence the selection of the next songs. The key idea here is to use {\em dynamic preference aggregation} to maintain fairness, and ensure that all users in a group are eventually satisfied.

Moreover, simply measuring how long a player has not been served might be an oversimplistic approach in some situations. Some players might have more pressing needs than others, or might be more critical than others. Consider, for example, a game where the EM is trying to maximize engagement in order to prevent players from leaving the game. Players that are not likely to quit the game would be less critical and should have a lower weight in the aggregation. Another example is that an EM can give more weight to the influencers in a game so that this group can propagate the desired effect to the rest of the players. In summary, general techniques to dynamically address preference aggregation over time is an open challenge and requires further research.

\subsubsection{Dynamic Grouping}\label{sec:groupmodels}

\begin{figure}[tb]
    \centering
    \includegraphics[width=\columnwidth]{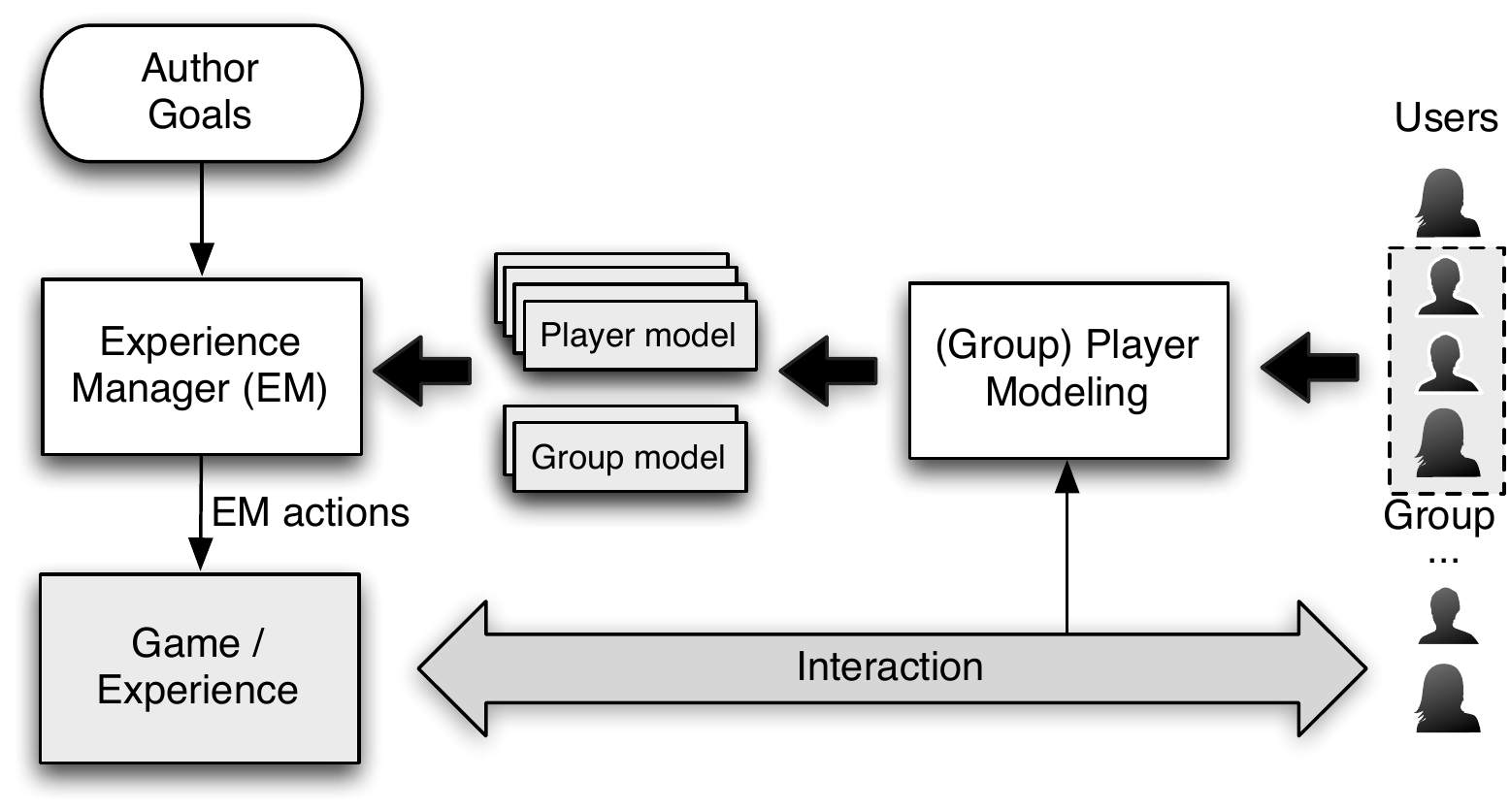}
    \caption{In a multi-player EM setting, instead of a single player, player modeling needs to model multiple players, as well as potentially nested groups that appear among the players. }
    \label{fig:mpem}
\end{figure}


Existing work in group modeling or on modeling individual in a group treats individuals at the same level of abstraction. However, since computer games can include more complex group dynamics than, say, music selection, they are a rich domain to explore dynamic hierarchical grouping of players. Thus, as illustrated in Figure \ref{fig:mpem}, in a multi-player EM setting, individual player models might co-exist with player group models at different levels of abstraction (i.e., maybe a subgroup is part of a larger group). 


Imagine an EM designed to infer player goals, so that appropriate hints or impediments can be triggered, depending on the desired difficulty level by the game designer. Imagine this EM operates in an MMO where a set of players can form {\em guilds} for missions, e.g., to gather the necessary ingredients to craft an advanced item. The purpose of the EM is to infer the goals of the players and provide appropriate hints or impediment. If the EM looks at each player individually or all the players in the game as a whole, it would be hard to capture what this particular set of players is trying to accomplish. 


Although player, and user modeling in general, has been studied for decades, approaches for performing group or player modeling are not as common\footnote{Notice that there is work on player models that apply to populations or ``classes'' of players, such as the idea of ``player types'' or heat maps (see the review of Smith et al. \cite{smith2011inclusive} for examples). However, those still focus on modeling individual players, using data from large populations.
}. The ability for EM to dynamically detect and assemble hierarchical groupings of players is thus another important direction.




\subsection{EM-driven Procedural Content Generation}\label{sec:pcg}


Procedural content generation (PCG)~\cite{shaker2016procedural} studies the algorithmic generation of content such as levels, music, and stories. One challenge in PCG is the issue of {\em controllability}, i.e., how to guide PCG algorithms to create content that satisfies certain conditions of interest. For example, how to design a map generation system that would generate levels with a desired level of difficulty. PCG is a promising idea for experience management since it would allow the EM to adapt the game/experience by generating new content. This idea has been explored in the context of both story generation~\cite{riedl2008toward}, and level generation for educational games~\cite{valls2017graph}.

Recent work in the PCG community has started to look at the problem of integrating player/user models into PCG algorithms to generate customized content~\cite{snodgrass2016controllable,shaker2010towards}. Multi-player EM, however, poses a harder challenge: that of generating content to suit groups of players, taking into account all the issues mentioned above (preference aggregation, fairness, or even social dynamics). Moreover, not even generation but just understanding how different types of generated content would affect player behavior, let alone their social dynamics and other multi-player aspects, is still an open problem.

\subsection{Evaluation Methods}

EM-based adaptive multi-player games represent a unique challenge in terms of how to evaluate their effectiveness. The main problem being the complexity of the EM components, compounded by human players' group behavior. The A/B testing methodology, both between-subject design and within-subject design, widely used in single-player EM with a relatively small number of players is inadequate for two reasons. First, human players' group dynamics can differ broadly based on context. Even the same group of players can play the same game very differently for a second time. This difference may be captured by player modeling and amplified through adaptation. In order to make a meaningful comparison between the EM-driven adaptive game and a baseline, it requires the subjects of the user study to go through comparable experiences. To account for the different gameplay experience due to the adaptive features, evaluating multi-player EM will require more data points. 

Second and related, currently widely used single EM evaluation methods treat the whole EM as a blackbox. For example, PaSSAGE~\cite{thue2008passage} evaluates its EM by comparing the player experiences 1) when the EM decides the next gameplay element, and 2) when the system chooses the next gameplay element. While this offers some insights into how well the EM works in conjunction with the game, this methodology cannot shed light onto how well different components of EM work. Since multi-player EM is more complicated than its single-player equivalent, the all-in-all evaluation approach is inadequate to guide further improvements since we do not know which component under-performed. Following the above MMO hint adaptation example. If a guild cannot complete its quest and reports low satisfaction, the current evaluation method cannot pinpoint the problem. Is it the individual player modeling that is not working correctly? Is the question how it aggregates the different models? Or did EM choose to cater to the wrong player in the wrong moment? To push the front of multi-player EM, the research community needs to develop more an in-depth and precise evaluation methodology.

\section{Conclusions}\label{sec:conclusions}


In this paper, we focused on Experience Management (EM) in the context of multi-player games. We identify both potentially impactful applications as well as a set of critical open research challenges to drive this area significantly forward.

Specifically, we argue that some of the key research challenges that need to be addressed to realize multi-player EM systems are: 
(1) player model aggregation: understanding how to aggregate player models from multiple players in order to make EM decisions about groups, (2) dynamic aggregation and grouping over time: how do we ensure the EM caters to the right players over time for fairness and effectiveness; (3) EM-driven PCG methods: design new procedural content generation algorithms that can generate content customized not to one player, but to a specific group of players; and (4) Evaluation methods: how to design new methodologies to evaluate multi-player EM approaches accounting for the additional issues caused by having groups of players and more complex EM approaches.
%
We believe that addressing these problems can open up a number of novel types of multi-player adaptive games and experiences and expand our knowledge of how to build adaptive experiences.


{\em Acknowledgements}: this work was partially funded by NSF award \#1816470. We would also like to thank the anonymous reviewers for the multiple suggestions and ideas that helped us improve the final version of this paper.

\bibliographystyle{IEEEtran}
\bibliography{references}


\end{document}